\documentclass[prb,twocolumn,showpacs,superscriptaddress]{revtex4}

\usepackage{graphicx}
\usepackage{epsfig}
\usepackage{verbatim}
\usepackage{amssymb}
\usepackage{amsfonts,amsmath}
\usepackage{bm}
\usepackage[T1]{fontenc}
\usepackage{eufrak}
\usepackage[colorlinks]{hyperref}
\hypersetup{colorlinks=true,
	linkcolor=red,
	anchorcolor=green,
	citecolor=blue,
	urlcolor=green
}

\providecommand{\be}{\begin{equation}}
\providecommand{\ee}{\end{equation}}
\providecommand{\bea}{\begin{eqnarray}}
\providecommand{\eea}{\end{eqnarray}}
\providecommand{\beas}{\begin{eqnarray*}}
\providecommand{\eeas}{\end{eqnarray*}}

\providecommand{\beni}{\begin{equation*}}
\providecommand{\eeni}{\end{equation*}}

\providecommand{\bw}{\begin{widetext}}
\providecommand{\ew}{\end{widetext}}

\newcommand{\nn}{\nonumber}

\newlength{\bilderlength}

\newlength{\figsize}
\begin{document}

 \title{Role of the Tracy-Widom distribution in the finite-size fluctuations of the critical temperature of the Sherrington-Kirkpatrick spin glass}
\author{Michele Castellana}
\email{michele.castellana@lptms.u-psud.fr}
\affiliation{LPTMS, CNRS and Universit\'{e} Paris-Sud, UMR8626, B\^{a}t. 100, 91405 Orsay, France} 
\affiliation{Dipartimento di Fisica, Universit\`a di Roma `La Sapienza' , 00185 Rome, Italy}
\author{Elia Zarinelli}
\email{elia.zarinelli@lptms.u-psud.fr}
\affiliation{LPTMS, CNRS and Universit\'{e} Paris-Sud, UMR8626, B\^{a}t. 100, 91405 Orsay, France} 

\pacs{64.70.Q-,75.10.Nr,02.50.-r }

\begin{abstract} 
We investigate the finite-size fluctuations due to quenched disorder of the
critical temperature of the Sherrington-Kirkpatrick spin glass. In order to accomplish this task, we perform a finite-size analysis of the spectrum of the susceptibility matrix obtained via the Plefka expansion. By exploiting results from random matrix theory, we obtain that the fluctuations of the critical temperature are described by the Tracy-Widom distribution with a non-trivial scaling exponent $2/3$.  
\end{abstract}

\maketitle

\section{Introduction} \label{intro}

The characterization of phase transitions in terms of  a non-analytic behavior of  thermodynamic functions in the infinite-size limit has served as a milestone\cite{lee1952statistical,yang1952statistical, huang, biskup2000general, ruelle1971extension} in the physical understanding of critical phenomena. In laboratory and numerical experiments the system size is always finite, so that  the divergences that would result from such a non-analytical behavior are suppressed, and  are replaced by smooth maxima occurring in the observation of physical quantities as a function of the temperature. In disordered systems the pseudo-critical temperature, defined as the temperature at which this maximum occurs, is a fluctuating quantity depending on the realization of the disorder. A question naturally  arises: can the fluctuations of the pseudo-critical temperature be understood and determined with tools of probability theory? Several efforts have been made to study the fluctuations of the pseudo-critical temperature for disordered finite-dimensional systems\cite{monthus2005delocalization,monthus2005distribution,igloi2007finite,monthus2006freezing} and their physical implications. For instance, recently Sarlat et al.\cite{sarlat2009predictive} showed that the  theory of finite-size scaling, which is valid for pure systems, fails in a fully-connected disordered models because of strong sample-to-sample fluctuations of the critical temperature.
\\

The Extreme Value Statistics of independent random variables is a well-established problem with a long history dating from the original work of Gumbel\cite{gumbel1958statistics}, while less results are known in the case where the random variables are correlated. The eigenvalues of a Gaussian random matrix are an example of strongly-correlated random variables\cite{mehta2004random}. Only recently, Tracy and Widom calculated\cite{tracy2002proceeding,tracy1996orthogonal,tracy1994level,tracy1993level} exactly the probability distribution of the typical fluctuations of the largest eigenvalue of a Gaussian random matrix around its mean value. This distribution, known as Tracy-Widom distribution, appears in many different models of statistical physics, such as directed polymers\cite{johansson2000shape,baik2000limiting} or polynuclear growth models\cite{prahofer2000universal}, showing profound links between such  different systems. 
 Conversely, to our knowledge no evident connections between the Tracy-Widom distribution and the physics of spin glasses have been found heretofore\cite{biroli2007extreme}.       
\\

The purpose of this work is to try to fill this gap. We consider  a mean-field spin glass model, the Sherrington-Kirkpatrick (SK) model\cite{sherrington1975solvable}, and propose a definition of finite-size critical temperature inspired by a previous analysis\cite{igloi2007finite}. We investigate the finite-size fluctuations of this pseudo-critical temperature  in the framework of Extreme Value Statistics and show that the  Tracy-Widom distribution naturally arises in the description of such fluctuations.


\section{The model}
 
The SK model\cite{sherrington1975solvable} is  defined by the Hamiltonian
\begin{equation} \label{3}
H [ \{ S_i\} ,\{ x_{ij} \}]= - \frac{J}{N^{1/2}} \sum_{i > j= 1}^ {N}x_{ij}S_i S_j  + \sum_{i=1}^ {N} h_i S_i
\end{equation}
where $S_i = \pm 1$, the couplings $\{ x_{ij} \}_{i > j=1, \cdots, N} \equiv \{ x \},\, x_{ji} \equiv x_{ij} \forall i>j$ are distributed according to normal distribution with zero mean and unit variance
\be \label{76}
P(x) = \frac{1}{\sqrt{2 \pi}}\textrm{e}^ {-\frac{x^2}{2}},
\ee
and $J$ is a parameter tuning the strength of the interaction energy between spins. \\
 
 The low-temperature features of the  SK model have been widely investigated in the past and are encoded in Parisi's solution\cite{parisi1980order,parisi1983order, talagrand2003generalized, MPV,0,NishimoriBook01}, showing that the SK has a finite-temperature spin glass transition at $T_c = J$ in the thermodynamic limit $N \rightarrow \infty$. The critical value $T_c$ can be physically thought as the value of the temperature where ergodicity breaking occurs and the spin glass susceptibility diverges\cite{MPV,NishimoriBook01,0}. \\
 
While Parisi's solution has been  derived within the replica method framework,  an alternative approach to study the SK model had been previously proposed by Thouless, Anderson and Palmer (TAP)\cite{thouless1977solution}. Within this  approach, the system is described in terms of a free-energy at fixed local magnetization, and  the physical features derived in terms of the resulting free-energy landscape. Later on,  Plefka\cite{plefka1982convergence} showed that the TAP free-energy can be obtained as the result of a systematic expansion in powers of the parameter 
 \[
 \alpha \equiv \frac{\beta J} { N^{1/2}},
\] 
 where $\beta$ is the inverse temperature of the model.  This $\alpha$-expansion, known as Plefka expansion, has thus served as a method for deriving TAP free energy for several class of models, and has been extensively used in several different contexts in physics, from classical disordered systems\cite{georges1990low,yedidia1990fully,yokota1995ordered}, to general quantum systems\cite{plefka2006expansion, ishii1985effect, de1992cavity, biroli2001quantum}. It is a general fact that, if the model is defined on a complete graph, the Plefka expansion truncates to a finite order in $\alpha$, because higher-order terms  should vanish in the thermodynamic limit. In particular, for the SK model the orders of the expansion larger than three are   believed\cite{yedidia2001idiosyncratic}  to vanish in the limit $N \to \infty$, in such a way that the expansion truncates, and  one is left with the first three orders of the $\alpha$-series, which read 
\begin{eqnarray} \label{3b}
& & -\beta f(\{ m_i \}, \beta) =  \nonumber \\ & &   - \sum_{i} \left[ \frac{1+m_i}{2} \ln \left(  \frac{1+m_i}{2} \right) + \frac{1-m_i}{2} \ln \left(  \frac{1-m_i}{2} \right) \right] \nonumber \\ 
 & & +\alpha  \sum_{i>j} x_{ij} m_i m_j   \nonumber \\ 
  & &+  \frac{\alpha  ^2}{2} \sum_{i>j} x_{ij}^2(1-m_i^2) (1-m_j^2)  , 
 \label{eq:tapf} 
\end{eqnarray}
where  $m_i \equiv \left \langle S_i \right\rangle$ is the local magnetization, i. e. the thermal average $\left\langle \right\rangle$  of the spin $S_i$ performed with the Boltzmann weight given by Eq. (\ref{3}) at fixed disorder $\{ x \}$.\\

In the thermodynamic limit $N \to \infty$, for temperatures $T>T_c$ the only minimum of $\beta f(\{ m \},\beta)$ is the paramagnetic one  $m_i = 0$ $\forall i$. Below the critical temperature, the TAP free energy has exponentially-many different minima: the system is in a glassy phase. In this framework, the phase transition at $T_c$ can be characterized by the inverse susceptibility matrix, which is also the Hessian  of $f$
\be \label{73}
\beta \chi^ {-1}_{ij}  \equiv  \beta \frac{\partial h_i}{\partial m_j}  =  \frac{\partial^ 2 (\beta f)}{\partial m_i \partial m_j} \ .
\ee
The inverse susceptibility matrix in the paramagnetic minimum at leading order in $N$ is:
 \begin{equation}\label{eq:hes}
\beta \chi^ {-1}_{ij} = (1 + \beta^2 J^2) \delta_{ij} - \alpha x_{ij}  \quad .
\end{equation}
Random-matrix theory states that the average density of eigenvalues of  $x$
\be 
\label{75}
\rho_N(\lambda) \equiv  \mathbb{E}_{x} \left[ \frac{1}{N} \sum_{i=1}^ N \delta(\lambda -\lambda_i(\{ x \})) \right],
\ee
 has a semi-circular shape\cite{wigner1955characteristic} on a finite support $[-2\sqrt{N}, 2\sqrt{N} ]$, where $\mathbb{E}_{x}$ denotes expectation value with respect to the random bonds $\{ x \}$, and $\lambda_i(\{ x \}) $ is the $i$-th eigenvalue of $x$. 
Eq. (\ref{75}) is nothing but the density of eigenvalues of the  Gaussian Orthogonal Ensemble (\textrm{GOE})  of Gaussian random matrices\cite{mehta2004random, fyodorov2005introduction}. \\
 
Due to self-averaging properties, the  minimal eigenvalue of $\beta \chi^{-1}$ in the paramagnetic minimum is $\lambda  = (1- \beta J)^2$.  This shows that, for $T>T_c$, $\lambda $ is  strictly positive and vanishes at $T_c$, implying the divergence \cite{MPV} of the spin glass susceptibility $1/\beta^2 \textrm{Tr}[\chi^2]$. Since $\lambda$ is also the minimal eigenvalue of the Hessian matrix of $\beta f$ in the paramagnetic minimum, we deduce that this is stable for $T>T_c$ and becomes marginally stable at $T_c$.
\\

This analysis sheds some light on the nature of the spin glass transition of the SK model in terms of the minimal eigenvalue $\lambda$ of the inverse susceptibility matrix (Hessian matrix) in the thermodynamic limit. In this paper we are intended to generalize such analysis to finite-sizes, where no diverging susceptibility neither  uniquely-defined critical temperature exist, and the minimal eigenvalue $\lambda$ acquires fluctuations due to quenched disorder. We  show that a finite-size pseudo-critical temperature can be suitably defined and investigate its finite-size fluctuations with respect to disorder.  As a result of this work, these fluctuations  are  found to be described by the Tracy-Widom distribution.  
\\

The rest of the paper is structured as follows. In Section \ref{sec2}, we generalize Eq. (\ref{eq:hes}) to finite sizes, in the simplifying assumption that the Plefka expansion can be truncated up to order $\alpha^2$, which is known as the TAP approach. We then study the finite-size fluctuations of the minimal eigenvalue $\lambda$ of the susceptibility matrix, and show that they are governed by the TW distribution.  

In Section \ref{sec3},  we extend this simplified approach by taking into account the full Plefka expansion, by performing an infinite re-summation of the series. 

Hence, in Section \ref{sec4}, we give a suitable definition of a finite-size pseudo-critical temperature,  and show that its fluctuations are  governed by the TW distribution. In Section \ref{disc}, this result is discussed in the perspective of generalizing it to more realistic spin glass models. \\

\section{Finite-size analysis of the susceptibility in the TAP approximation}\label{sec2}

In this Section we study the finite-size fluctuations due to disorder of the minimal eigenvalue of the inverse susceptibility matrix $\beta \chi^ {-1}$  at the paramagnetic minimum $m_i = 0\,  \forall i$, by  considering the free energy $f$ in the TAP approximation, Eq. (\ref{3b}).
We want to stress the fact that large deviations of thermodynamics quantities of the SK model have been already studied heretofore. For example, Parisi et al. have studied\cite{PhysRevLett.101.117205, PhysRevB.79.134205} the probability distribution of large deviations of the free energy  within the replica approach. The same authors studied  the probability of positive large deviations of the free energy per spin in general mean-field spin-glass models\cite{PhysRevB.81.094201}, and showed that such fluctuations can be interpreted in terms of the fluctuations of the largest eigenvalue of Gaussian matrices, in analogy with the lines followed in the present work. \\

Back to the TAP equations (\ref{3b}), the inverse susceptibility matrix in the paramagnetic minimum for finite $N$ reads:

\bea  \label{31} \nn 
\beta \chi^ {-1}_{ij} &  = &   - \alpha x_{ij} + \delta_{ij} \left(1 + \alpha^2 \sum_{k\neq i} x^ 2_{ki} \right) \\ 
 &=&- \alpha x_{ij} + \delta_{ij} \left( 1+ \beta^ 2 J^2 \right)   + \delta_{ij}  \frac{ (\beta J)^ 2}{\sqrt{N}} z^ i_2 ,
\eea 
where 
\bea \label{30}
z^ i_2 & \equiv  & \sqrt{N}\left( \frac{1}{N}  \sum_{k\neq i} x^ 2_{ki} - 1 \right) .
\eea

According to Eq. (\ref{30}),  $z_2^ i $ is given by the sum of $N-1$ independent identically-distributed  random variables $x^ 2_{ij}$. By the central limit theorem, at leading order in $N$ the variable $z_2^i$ is distributed according to a Gaussian distribution with zero  mean  and variance $2$
\be \label{71}
p_N(z_2^i=z) \overset{N \rightarrow \infty}{ \rightarrow} \frac{1}{\sqrt{4 \pi}} \textrm{e}^{-z^2/4},
\ee
where $p_N(z_2^i=z)$ denotes the probability that $z_2^i $ is equal to $z$ at finite size $N$. \\

We set
\be
\beta \chi^ {-1}_{ij} \equiv \delta_{ij} \left( 1+ \beta^ 2 J^2 \right) + \alpha M_{ij} \ \  .
\ee
According to Eq.  (\ref{30}), the diagonal elements of $M_{ij}$ are random variables correlated to out-of-diagonal elements.
 The statistical properties of the spectrum of a random matrix whose entries are correlated to each other has been studied heretofore only in some cases.  For instance, Staring et al.\cite{staring2003random} studied the mean eigenvalue density for matrices with a constraint implying that the row sum of matrix elements should vanish, and other correlated cases have been investigated both from a physical\cite{shukla2005random}  and  mathematical\cite{bai2008large}  point of view. 
\\

In recent years, a huge amount of results has been obtained on the distribution of the minimal eigenvalue of a $N \times N$ random matrix drawn from Gaussian ensembles, such as \textrm{GOE}. In particular, Tracy and Widom\cite{tracy2002proceeding,tracy1996orthogonal,tracy1994level,tracy1993level} deduced that for large $N$, small  fluctuations of the minimal eigenvalue $\lambda_{\textrm{GOE}}$ of a \textrm{GOE} matrix around its leading-order value $-2 \sqrt{N}$ are given by 
\be \label{25}
\lambda_{\textrm{GOE}} = -2 \sqrt{N} + \frac{1}{N^ {1/6}} \phi_{\textrm{GOE}} , 
\ee   
where  $\phi_{\textrm{GOE}}$ is a random variable distributed according to the Tracy-Widom (TW) distribution for the \textrm{GOE} ensemble $p_{\textrm{GOE}}(\phi)$.
 It follows that for $\beta J=1$ if $z_2^ i$ was independent on $\{ x \}$, the matrix $M_{ij}$ would belong to the  \textrm{GOE} ensemble, and
the minimal eigenvalue $\lambda$ of $\beta \chi^ {-1}$ would define a variable $\phi$ according to 
\be \label{26}
\lambda = \frac{1}{N^ {2/3}} \phi,
\ee
  and $\phi$ would be distributed according to the TW distribution $p_{\textrm{GOE}}(\phi)$. \\
  
As shown in Appendix \ref{app1}, this is indeed the case for $z_2^ i$, which can be treated, at leading order in $N$, as a random variable independent on $x_{ij}$. The general idea is that $z_2^ i $ is given by the sum of $N-1$ terms all of the same order of magnitude, and only one amongst these $N-1$ terms depends on $x_{ij}$. It follows that at leading order in $N$, $z_2^ i $ can be considered as independent on $x_{ij}$. Since in Eq. (\ref{31}) $z_2^ i$ is multiplied by a sub-leading factor $1/\sqrt{N}$, in Eq. (\ref{31}) we can consider $z_2^ i $ at leading order in $N$, and treat it as  independent on $x_{ij}$.\\

\begin{figure}[t]
\begin{centering}
\includegraphics[width=8cm]{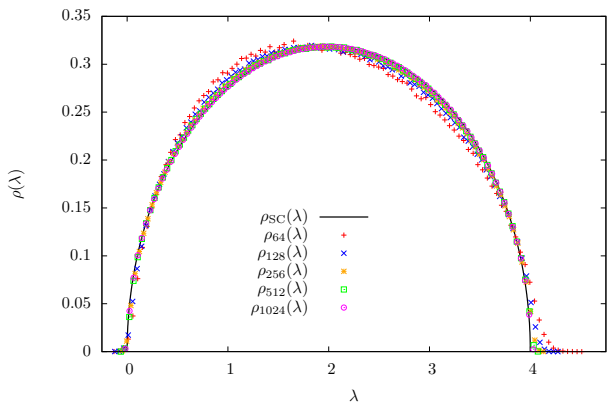}
\caption{Density of eigenvalues $\rho_N(\lambda)$ of the matrix $\beta \chi^ {-1}$  for $N=64,128,256,512,1024$ (in red, blue, yellow, green, violet respectively), $\beta J =1$ and $S=16\times 10^3$, and  Wigner semicircular law $\rho_{\textrm{SC}}(\lambda) = 1/(2 \pi) \sqrt{4-(2-\lambda)^2} $ (black) as a function of $\lambda$. $\rho_N(\lambda)$ approaches $\rho_{\textrm{SC}}(\lambda)$ as $N$ is increased.   }
\label{fig3}
\end{centering}
\end{figure}

To test this independence property, we set $\beta J = 1$, generate numerically $S \gg 1$ samples of the $N \times N$ matrix $\beta \chi^ {-1}$, and  compute the average density of eigenvalues  
of $\beta \chi^ {-1}$,  defined  as in Eq. (\ref{75}), together with the distribution of the minimal eigenvalue $\lambda$ for several sizes $N$. The eigenvalue distribution $\rho_N(\lambda)$ as a function of $\lambda$ is depicted in Fig. \ref{fig3}, and tends to the Wigner semicircle as $N$ is increased, showing that the minimal eigenvalue $\lambda$ tends to $0$ as $N \rightarrow \infty$. 
   
\begin{figure*}[t]
\begin{centering}
\includegraphics[width=14cm]{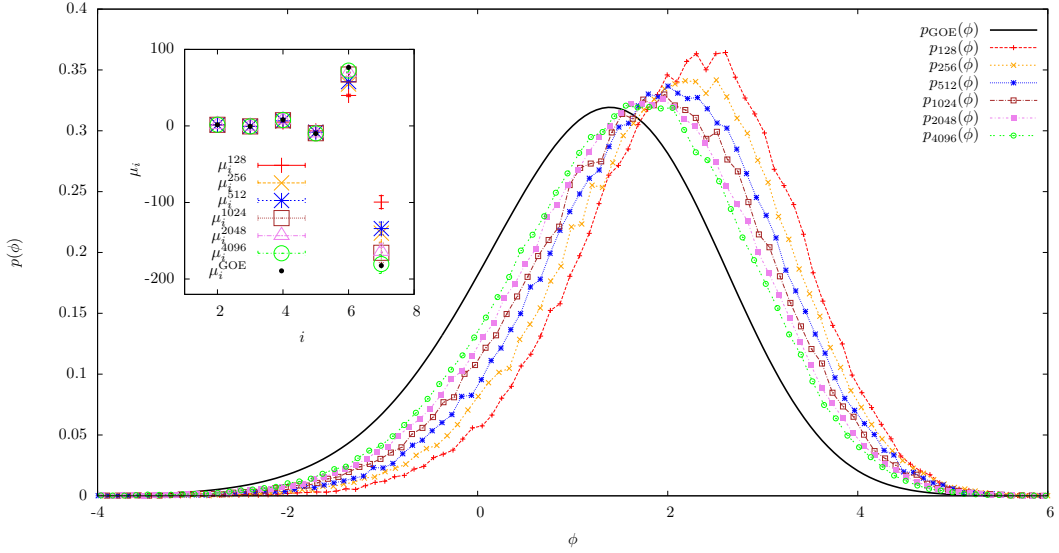}
\caption{Distribution $p_N(\phi)$  for $N=128, 256, 512, 1024,2048, 4096$ (in red, yellow, blue, brown, violet, green respectively) and $10^5 \leq S \leq 4 \times 10^5$ samples,  and the Tracy-Widom distribution $p_{\textrm{GOE}}(\phi)$ for the \textrm{GOE} ensemble (black), as a function of $\phi$. For increasing $N$, $p_N(\phi)$ approaches $p_{\textrm{GOE}}(\phi)$, confirming the asymptotic independence of the diagonal elements (\ref{25}) by each of the off-diagonal elements $x_{ij}$ for large $N$. Inset: $\mu^ N_i$ for sizes $N=128,256,512,1024,2048,4096$ (in red, yellow, blue, brown, violet, green respectively), $10^5 \leq S \leq 4 \times 10^5$, and $\mu_i^ {\textrm{GOE}}$ (black) as a function of $i>1$.    }
\label{fig1}
\end{centering}
\end{figure*}

   The finite-size fluctuations of $\lambda$ around $0$ are then investigated in Fig. \ref{fig1}. Defining $\phi$ in terms of $\lambda$ by Eq. (\ref{26}), in Fig. \ref{fig1} we depict the distribution $p_N(\phi)$ of the variable $\phi$ for several sizes $N$, and show that for increasing $N$, $p_N(\phi)$ approaches the TW distribution $p_{\textrm{GOE}}(\phi)$. 
Let us introduce  the central moments 
   \beas
   \mu^ N_1 & \equiv & \mathbb{E}_N[\phi ],\\
   \mu^ N_ i & \equiv &  \mathbb{E}_N[  (\phi - \mathbb{E}_N[\phi ])^ i ]\,\forall  i > 1
\eeas
of $p_N(\phi)$, and the central moments 
\beas
\mu ^ {\textrm{GOE}}_1 & \equiv & \mathbb{E}_{\textrm{GOE}}[\phi] ,\\
   \mu^ {\textrm{GOE}}_ i & \equiv &  \mathbb{E}_{\textrm{GOE}}[  (\phi - \mathbb{E}_{\textrm{GOE}} [\phi ])^ i ]\, \forall i>1
\eeas
of the TW distribution, where
\beas
\mathbb{E}_N[ \cdot ] &\equiv & \int d\phi \, p_{N} (\phi) \cdot ,\\ \nn
   \mathbb{E}_{\textrm{GOE}}[ \cdot ] &\equiv & \int d\phi \, p_{{\textrm{GOE}}} (\phi) \cdot. 
\eeas
In the inset of Fig. \ref{fig1} we depict $\mu^ N_i$ for several sizes $N$ and $\mu_i^ {\textrm{GOE}}$ as a function of $i$, showing that  $\mu^ N_i$ converges to $\mu^ {\textrm{GOE}}_i$ as $N$ is increased. 

\begin{figure}
\begin{centering}
\includegraphics[width=8cm]{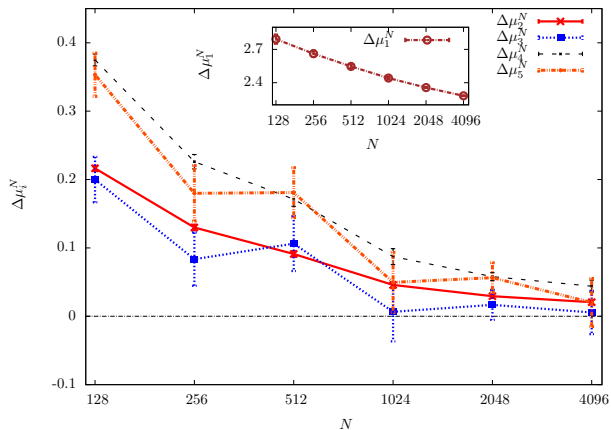}\hspace*{3cm}
\caption{   Relative difference $\Delta \mu ^ N_i $ between the central moments  $\mu^ {N}_ i$ of the distribution $p_N(\phi)$ for $10^5 \leq S \leq 4 \times 10^5$, and the central moments $\mu^ {\textrm{GOE}}_ i$ of the Tracy-Widom distribution as a function of $N=128, 256, 512, 1024,2048, 4096$,  for $i=2, 3, 4, 5$ (in red, blue, black, orange  respectively). For increasing $N$, $\mu^ N _ i$ approaches $\mu^ {\textrm{GOE}}_i $, confirming the asymptotic independence of $z_2^ i$ by each of the off-diagonal elements $x_{ij}$ for large $N$.
Inset: relative difference of the first central moment $\Delta \mu ^ N_1$ as a function of $ N$ (brown). $\Delta \mu ^ N_1$ approaches $0$ very slowly as $N$ is increased.
  }
\label{fig4}
\end{centering}
\end{figure}

In Figure \ref{fig4} this convergence is clarified by depicting $\Delta \mu^ N_i \equiv (\mu^ N_i - \mu_i^ {\textrm{GOE}} ) /  \mu_i^ {\textrm{GOE}}$ for several values of $i>1$  as a function of $N$. $\Delta \mu^ N_i$  is found to converge to $0$ for large $N$. In the inset of Fig. \ref{fig4} we depict $\Delta \mu^ N_1 $ as a function of $N$,  showing that the convergence of the first central moment with $N$ is much slower than that of the other central moments. It is interesting to observe that a slowly-converging first moment has been recently found also in experimental\cite{takeuchi2010universal} and numerical\cite{rambeau} data of models of growing interfaces where the TW distribution appears. \\

The analytical argument proving the independence property of $z_2^ i$ has been thus confirmed by this numerical calculation. Hence, the main result of this Section is that the finite-size fluctuations of the minimal eigenvalue of the susceptibility matrix $\beta \chi^ {-1}$ in the TAP approximation  for $\beta J=1$ are of order $N^ {-2/3}$ and are distributed according to the TW law. These fluctuations have already been found to be of order $N^ {-2/3}$ in a previous work\cite{bray1979evidence}, and more recently reconsidered\cite{aspelmeier2008finite}, following an independent derivation based on scaling arguments, even though the distribution has not been worked out. Our approach sheds some light on the nature of the scaling $N^ {-2/3}$, which is non-trivial, since it comes from the $N^{-1/6}$-scaling of the TW distribution, which is found to govern the fluctuations of $\lambda$. Moreover, the fact that we find the same scaling as those found in such previous works can be considered as a consistency test of our calculation.\\
 
 We now recall that  both the derivation of this Section and  the  previously-developed analysis of  Bray and Moore\cite{bray1979evidence} rely on the TAP approximation, i. e.   neglect
 the terms of the Plefka expansion (\ref{1}) of order larger than $2$
 in $\alpha$.  As we will show in the following Section, these terms
 give a non-negligible  contribution to the finite-size corrections of the TAP equations, and so to the finite-size fluctuations of the critical temperature, and thus must be definitely taken into account in a complete treatment. \\

   \section{Finite-size analysis of the susceptibility within the full Plefka expansion} \label{sec3}

   In this Section we  compute the inverse susceptibility matrix $\beta \chi^{-1}$ by taking into account all  the terms of the Plefka expansion, in the effort to go beyond the TAP approximation of Section  \ref{sec2}.    Notwithstanding its apparent difficulty, here we show that this task can be pursued   by a direct inspection of the terms of the expansion. Indeed, let us formally write the free-energy $f$ a a series\cite{plefka1982convergence} in $\alpha$, 
\be \label{1}
f(\{ m \} , \beta) = \sum_{n=0}^ {\infty} \alpha^ n f_n(\{ m \}, \beta ).
\ee

 For $n<3$, the $f_n$s  are given by Eq. (\ref{3b}). For   $n>3$, $f_n$ is given by the sum of several different addends\cite{yedidia2001idiosyncratic}, which proliferate for increasing $n$.
   It is easy to show that at leading order in $N$, there is just one term contributing to $f_n$, and that such  term can be written explicitly as
\bea \label{2}
f_n (\{ m \}, \beta ) &  \overset{N \rightarrow \infty }{\approx}   & \sum_{i_1 > \cdots > i_{n-1}} x_{i_1 i_2} x_{i_2 i_3} \cdots x_{i_{n-1} i_1}  \\ \nn
&& \times(1-m_{i_1}^2) \times \cdots \times (1-m_{i_{n-1}}^2).
\eea 
It follows that by plugging Eq. (\ref{2}) in Eq. (\ref{1}) and computing $\beta \chi^ {-1}$ for $m_i=0$, one obtains a simple expression for the inverse susceptibility at the paramagnetic solution 
\bw
\bea \label{6} \nn 
\beta \chi^ {-1}_{ij} & = &  - \alpha x_{ij} + \delta_{ij} \left(1 + \alpha^2 \sum_{k\neq i} x^ 2_{ki} + 2 \sum_{n=3}^ {\infty} \alpha^ n  \sum_{i_1 > \cdots > i_{n-1}} x_{i i_1} x_{i_1 i_2} \cdots x_{i_{n-1} i}  \right)\\ 
& = & - \alpha x_{ij} + \delta_{ij} \left( 1+ \beta^ 2 J^2 \right)   + \delta_{ij}  \frac{1}{\sqrt{N}} \left[ (\beta J)^2 z_2^ i + 2 \sum_{n=3}^ {\infty} \frac{ (\beta J)^ n}{\sqrt{(n-1)!}} z^ i_n \right] . 
\eea
\ew
where 
\bea \label{5} \nn 
z^ i_n & \equiv  & \frac{\sqrt{(n-1)!}}{N^ {\frac{n-1}{2}}} \times \\ 
&& \times \sum_{i_1 > \cdots > i_{n-1}} x_{i i_1} x_{i_1 i_2} \cdots x_{i_{n-1} i},\, \forall n>2. 
\eea
According to Eq. (\ref{5}),  one has that at leading order in $N$
\bea\label{60} \nn 
\mathbb{E}_x[z^ i_n] &=& 0 \, \forall n>2,\\
     \mathbb{E}_x[(z^ i_n)^2]  &=& 1\,  \forall n>2, 
\eea
where in the second line of Eq. (\ref{60}) the multiple sum defining $z_n^ i$ has been evaluated at leading order in $N$. \\

%

We observe that the random variables $z^ i_n$ and $x_{jk}$ in Eq. (\ref{6}) are not independent, since each $z^ i_n$ depends on the bond variables $\{ x \}$. Following an argument similar to that  given in Section \ref{sec2} for $z_2^ i$, we observe that, by Eq. (\ref{5}) and at leading order in $N$,  $z^ i_n$ is given by a sum of $O(N^ {n-1})$ terms which are all of the same order of magnitude.  Each term is given by the product of $n-1$ bond variables $x_{i i_1} x_{i_1 i_2} \cdots x_{i_{n-1} i}$ forming a loop passing by site $i$. 
 For any fixed $i,j,k$ and $n$, only $O(N^ {n-2})$  terms amongst the $O(N^ {n-1})$ terms of $z^ i_n$ are entangled with the random bond variable $x_{jk}$. 
  It follows that at leading order in $N$, $z_n^ i$ can be considered as independent by $x_{jk}$. Since the sum in the second line of Eq. (\ref{6}) has a $1/\sqrt{N}$ factor multiplying each of the $z_n^ i$s, we can consider the $z_n^ i$ at leading order in $N$. Hence, in Eq. (\ref{6}) we  can consider each of $z_n^ i$s as independent on $x_{jk}$.  \smallskip  \\

In Appendix \ref{app2} we show that at leading order in $N$ the
distribution of $z^ i_n$ is a Gaussian with zero mean and unit
variance for every $i$ and $n>2$, while in Appendix \ref{app3} we show that at leading order in $N$ the
variables $\{ z_n^i \}_{n,i}$ are mutually independent. Both these predictions are confirmed by numerical tests, illustrated in Appendix \ref{app2} and \ref{app3} respectively.  \\

 Hence, at leading order in $N$ the  term in square brackets in Eq. (\ref{6}) is nothing but the sum of independent Gaussian variables, and is thus equal to a random variable $\sigma \times \zeta_i$, where $\zeta_i$ is Gaussian with  zero mean and unit variance, and
\beas
\sigma^2 & = & 2 (\beta J)^ 4 + 4 \sum_{n=3}^ {\infty} \frac{ (\beta J)^ {2n}}{(n-1)!}\\ \nn 
& = & 2 (\beta J)^2 \{ 2 (e^{(\beta J)^2}-1)-(\beta J)^2 \}
\eeas

 It follows that Eq. (\ref{6}) becomes 
\bea \label{7}\nn 
\beta \chi_{ij}^ {-1}  & = & - \alpha x_{ij} + \delta_{ij} \left( 1+ \beta^ 2 J^2 + \frac{\sigma}{\sqrt N} \zeta_i\right)   \\
 & = &  - \alpha x'_{ij} + \delta_{ij} \left( 1+ \beta^ 2 J^2 \right),
\eea 
where 
\be \label{8}
 x'_{ij} \equiv  x_{ij} - \delta_{ij}  \frac{\sigma}{\beta J} \zeta_i   . 
\ee

Because of the additional diagonal term  in Eq.  (\ref{8}), the matrix $x'_{ij}$ does not belong to the \textrm{GOE} ensemble.
Notwithstanding this fact, it has been shown by Soshnikov\cite{soshnikov1999universality} that the presence of the diagonal elements in Eq. (\ref{8}) does not alter the universal distribution of the maximal  eigenvalue of $x'_{ij}$, which is still distributed according to the TW law. Hence, denoting by $\lambda$  the minimal eigenvalue of $\beta \chi^ {-1}$, we have
\be \label{12}
\lambda = (1 - \beta J)^2 + \frac{\beta J}{N ^ {2/3}}\phi_{\textrm{GOE}},
\ee 
where $\phi_{\textrm{GOE}}$ is a random variable depending on the sample $x_{ij}$, and distributed according to the TW law. 
\\
%
\medskip\\

In this Section we have calculated the inverse susceptibility matrix $\beta \chi^{-1}$, by considering  the full Plefka expansion. In this framework  additional diagonal terms are generated that were not present in the TAP approximation. These additional terms can be handled via a resummation to all orders in the Plefka expansion. As a result, we obtain that the fluctuations of the  minimal eigenvalue $\lambda$ of the susceptibility $\beta \chi^ {-1}$ are still governed  by the TW law, as in the TAP case treated in Section \ref{sec2}.  \\


   \section{Finite size fluctuations of the critical temperature}\label{sec4}

We can now define a finite-size critical temperature, and investigate its finite-size fluctuations due to disorder.\\

In the previous Sections we have shown that for a large but finite size $N$, the minimal eigenvalue of the inverse susceptibility matrix, i. e.  the Hessian matrix of $\beta f(\{ m \} , \beta)$ evaluated in the paramagnetic minimum $m_i = 0$, is a function of the temperature and of a quantity $\phi_{\textrm{GOE}}$, which depends on the realization of the disorder $\{ x \}$. Since the TW law, i. e. the distribution of $\phi_{\textrm{GOE}}$, has support for both positive and negative values of $\phi_{\textrm{GOE}}$, the subleading term in Eq. (\ref{12}) can be positive or negative.   Accordingly, for  samples $\{ x \}$ such that $\phi_{\textrm{GOE}} <0$, there exists a value of $\beta J \approx 1$ such that $\lambda(\beta J)=0$, in such a way that  the spin-glass susceptibility in the paramagnetic minimum diverges. This fact is physically meaningless, since there cannot be divergences in physical quantities at finite size. 
This apparent contradiction can be easily understood by observing that if $\lambda(\beta J)=0$, the true physical susceptibility is no more the paramagnetic one, but must be evaluated  in the low-lying non-paramagnetic minima of the free-energy, whose appearance is driven by the emergent instability of the paramagnetic minimum. 
%
\\

According to this discussion, in the following we will consider only samples $\{ x \}$ such that $\phi_{\textrm{GOE}}>0$. 
For these samples, the spectrum of the Hessian matrix at the paramagnetic minimum has positive support for every temperature: the paramagnetic solution is  always stable and the paramagnetic susceptibility matrix $\chi$ is physical and finite. We define a pseudo-inverse critical temperature $\beta_c J$
as the value of $\beta J$ such that $\lambda$ has a minimum at $\beta_c J$
\bea \label{16}\nn
\left. \frac{d \lambda}{d \beta J}\right|_{\beta J = \beta_c J} & \equiv & 0\\  
& = & - 2 (1-\beta_c J)  + \frac{ 1}{N^ {2/3}} \phi_{\textrm{GOE}} 
\eea
where in the second line of  Eq. (\ref{16}), Eq. (\ref{12}) has been used. 
This definition of pseudo-critical temperature has a clear physical interpretation: the stability of  the paramagnetic minimum, which is encoded into the spectrum of the Hessian matrix $\beta \chi^{-1}$, has a minimum at $\beta = \beta_c$.
According to Eq. (\ref{16}), the finite-size critical temperature $\beta_c$ is given by 
\be \label{22}
\beta_c J = 1 - \frac{1/2}{N^ {2/3}} \phi_{\textrm{GOE}},
\ee
where $\phi_{\textrm{GOE}}$ depends on the sample $\{ x \}$, and is distributed according to the TW law. \medskip\\

Eq. (\ref{22}) shows that the pseudo-critical temperature of the SK model is a random variable depending on the realization of the quenched  disorder:  finite-size fluctuations of the pseudo-critical temperature are of order $N^ {-2/3}$, and are distributed according to the TW law. This has to be considered the main result of this paper. \medskip\\

\section{Discussion and conclusions}\label{disc}

In this paper, the finite-size fluctuations of the critical temperature of the Sherrington-Kirkpatrick spin glass model have been investigated. The analysis is carried on within the framework of the Plefka expansion for the free-energy at fixed local magnetization. A direct investigation of the expansion shows that an infinite resummation of the series is required to describe the finite-size fluctuations of the critical temperature. By observing that the terms in the expansion can be treated as independent random variables, one can suitably define a finite-size critical temperature.  
Such a critical temperature has a unique value in the infinite-size limit, while exhibits fluctuations due to quenched disorder at finite sizes. 
These fluctuations  with respect to the infinite-size value have been analyzed, and have been  found to be of order $N^{-2/3}$, where $N$ is the system size, and to be distributed according to the Tracy-Widom distribution.\\
 An analogous role of the TW distribution in the description of the critical properties of a physical system has also recently been clarified by Forrester et al.\cite{forrester2011non},  showing that the TW law describes the finite-size corrections of the free-energy of a Yang-Mills theory in the neighborhood of its critical point. \\

The exponent $2/3$ describing the fluctuations of the pseudo-critical temperature stems from the fact that the finite-size fluctuations of the  minimal eigenvalue $\lambda$ of the inverse susceptibility matrix are of order $N^{-2/3}$. Such a scaling for $\lambda$ at the critical temperature had already been obtained in a previous work\cite{bray1979evidence}, where it was derived by a completely independent method, by taking into account only the first three terms of the Plefka expansion. The present work  shows that a more careful treatment, including an infinite resummation of the expansion, is needed to handle finite-size effects. The exponent $2/3$ derived by Bray and Moore\cite{bray1979evidence} is here rederived by establishing a connection with recently-developed results in random matrix theory, showing that the scaling $N^ {-2/3}$ comes from the scaling of the Tracy-Widom distribution, which was still unknown when the paper by Bray and Moore\cite{bray1979evidence} had been written.\\

As a possible development of the present work, it would be interesting to study the fluctuations of the critical temperature for a SK model where the couplings are distributed according to a power-law. Indeed, in a recent work\cite{biroli2007top} the distribution of the  largest eigenvalue $\lambda$ of a random matrix $M$ whose entries $M_{ij}$ are power-law distributed as $p(M_{ij}) \sim M_{ij}^{-1- \mu}$ has been studied. The authors show that if $\mu>4$ the fluctuations of $\lambda$ are of order $N^{-2/3}$ and are given by the TW distribution, while  if $\mu<4$ the fluctuations are of order $N^{-2/\mu -1/2}$ and are governed by Fr\'echet's statistics. This result could be directly applied to a SK model with power-law distributed couplings. In particular, it would be interesting to see if there exists a threshold in the exponent $\mu$ separating two different regimes of the fluctuations of $T_c$. \\

Another  interesting perspective would be to generalize the present approach to realistic spin glass models with finite-range interactions. For instance, a huge amount of results has been quite recently obtained  for the three-dimensional Ising spin glass
\cite{banos2010nature, hasenbusch2008critical, alvarez2010static, belletti2009depth, contucci2007ultrametricity, contucci2009structure, krzakala2000spin,PhysRevB.58.14852}, 
and for the short-range $p$-spin glass model in three dimensions\cite{PhysRevB.58.12081},
yielding evidence for a finite-temperature phase transition. It would be interesting to try to generalize the present work to that systems, and compare the resulting fluctuations of the critical temperature with sample-to-sample fluctuations observed in these numerical works.  Accordingly, the finite-size fluctuations deriving from the generalization of this work to the three-dimensional Ising spin glass could be hopefully compared with those observed in experimental spin glasses\cite{gunnarson1991static}, such as $\textrm{Fe}_{0.5} \textrm{Mn}_{0.5} \textrm{TiO} _3$.\\

Finally,  a recent numerical analysis \cite{billoire2011finite} inspired by the present work has investigated the sample-to-sample fluctuations of a given pseudo-critical temperature for the SK model, which is different from that defined in this work. Even though the relatively small number of samples did not allow for a precise determination of the probability distribution of that pseudo-critical point, the analysis  yields a scaling exponent equal to $1/3$, which is different from that of the pseudo-critical temperature defined here. As a consequence, the general scaling features of the pseudo-critical temperature seem to depend on the actual definition of the pseudo-critical point itself, even though different definitions of the pseudo-critical temperature must all converge to the infinite-size pseudo-critical temperature as the system size tends to infinity. As a future perspective, it would be interesting to investigate which amongst the features of the pseudo-critical point are definition-independent, if any. 

\section*{Acknowledgements}

We are glad to thank J. Rambeau and G. Schehr for interesting discussions and suggestions. We also acknowledge support from the D. I. computational center of University \textit{Paris Sud}.

\appendix

\section{Proof of the asymptotic independence of $x_{ij}$ and $z_2^ i $} \label{app1}

Here we show that at leading order in $N$ the variables $x_{ij}$ and $z_2^ i $ are independent, i. e. that at leading order in $N$
\be  \label{40}
 p_N(x_{ij}  =  x, z_{2}^ i = z) =    p_N(x_{ij} = x) \times p_N(z_{2}^ i = z)  . 
\ee 
Let us explicitly write the left-hand size of Eq. (\ref{40}) as
\bw 
\bea \label{41}\nn 
 p_N(x_{ij}  =  x, z_{2}^ i = z) &=& \mathbb{E}_{\{x_{ik}\}_{k \neq i}} [ \delta(x_{ij} - x)  \delta (z_2^ i - z)], \\
  & = &  \mathbb{E}_{x_{ij}}\left[   \delta(x_{ij} - x)  \mathbb{E}_{\{ x_{ik} \}_{k \neq i, k \neq j}} \left[  \delta\left( \sqrt{N} \left( \frac{1}{N} \sum_{k\neq i, k\neq j} x_{ki}^ 2 -1 \right) - \overline{z}^ {ij}_2 \right) \right] \right] 
 \eea
 \ew

\begin{figure}
\begin{centering}
\includegraphics[width=8.5cm]{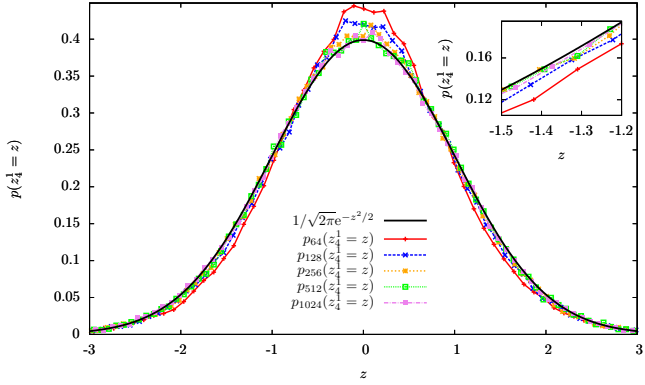}
\caption{   Probability distribution $p_{N}(z_4^1 = z)$ for
 $S=10^5$
 and different
  values of $N=64,128,256,512,1024$ (in red, blue, yellow, green, violet respectively) together with a Gaussian
  distribution $1/\sqrt{2\pi} \textrm{e}^{-z^2/2}$ with zero mean and
  unit variance (black), as a function of $z$. As $N$ is increased $p_{N}(z_4^1 = z)$ converges to $1/\sqrt{2\pi} \textrm{e}^{-z^2/2}$, as predicted by the analytical calculation, Eq. (\ref{p_of_z}). 
Inset: zoom of the above plot explicitly showing the convergence of $p_{N}(z_4^1 = z)$  to $1/\sqrt{2\pi} \textrm{e}^{-z^2/2}$ as $N$ is increased.}
\label{fig_test_z}
\end{centering}
\end{figure}

where $\mathbb{E}_{ x_{lm}, x_{no}, \cdots  }$ denotes the expectation value with respect to the probability distributions of the variables $  x_{lm} , x_{no}, \cdots $, $\delta$ denotes the Dirac delta function, and 
\be \label{43}
 \overline{z}^ {ij}_2 \equiv z - \frac{x_{ij}^2}{\sqrt{N}}.
\ee
Proceeding systematically at leading order in $N$, the  second expectation value in the second line of  Eq. (\ref{41}) is nothing but the probability that the variable $\sqrt{N} ( \frac{1}{N} \sum_{k\neq i, k\neq j} x_{ki}^ 2 -1 )$ is equal to $ \overline{z}^ {ij}_2$. We observe that according to the central limit theorem, at leading order in $N$ this probability is given by  
\bea \label{42} \nn
 &&\mathbb{E}_{\{ x_{ik} \}_{k \neq i, k \neq j}} \left[  \delta\left( \sqrt{N} \left( \frac{1}{N} \sum_{k\neq i, k\neq j} x_{ki}^ 2 -1 \right) - \overline{z}^ {ij}_2 \right) \right]    = \\
&&  \frac{1}{\sqrt{4 \pi}} \textrm{e}^ { - \frac{\left(\overline{z}_2^ {ij}\right)^2}{4} }.
\eea
By plugging Eq. (\ref{42}) into Eq. (\ref{41}) and using Eq. (\ref{43}), one has 
\bea \label{44}\nn 
 p_N(x_{ij}  =  x, z_{2}^ i = z) &=& \frac{1}{\sqrt{4 \pi}} \int dx_{ij}   P(x_{ij})\delta(x_{ij} - x) \times \\ \nn 
&& \times\textrm{e} ^{  - \frac{\left(z- x^ 2_{ij}/\sqrt{N}\right)^2}{4} } \\ \nn 
 & = & P(x)  \frac{1}{\sqrt{4 \pi}}  \textrm{e} ^{  - \frac{\left(z- x^2/\sqrt{N}\right)^2}{4} }  \\
 & = & p_N(x_{ij} = x) \times p_N(z_2^ i = z), 
 \eea
where in the first line Eq. (\ref{44}) we explicitly wrote the expectation value with respect to $x_{ij}$ in terms of the probability distribution (\ref{76}), while in the third line  proceeded at leading order in $N$, and used Eq. (\ref{71}).

\begin{figure}[htb]
\begin{centering}
\vspace{1cm}
\includegraphics[width=8.2cm]{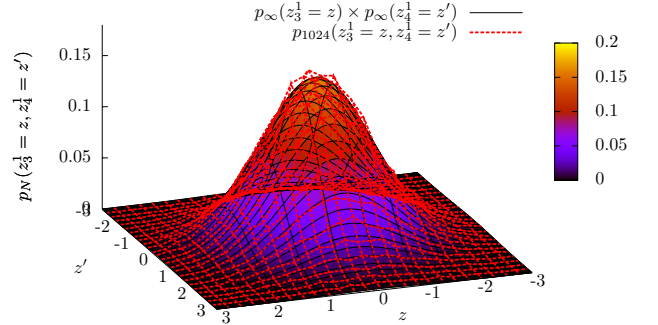}
\caption{$p_{1024}(z_3^1 = z, z_4^1=z')$ for 
$S=10^5$
samples (red), and  the $N \rightarrow \infty$-limit of  the right-hand side of Eq. (\ref{p_z_z}) (black), as a function of $z,z'$. For large $N$, $p_{N}(z_3^1 = z, z_4^1=z')$ equals $p_{N}(z_3^1 = z) \times p_{N}(z_4 ^1 = z') $, as predicted by Eq. (\ref{p_z_z}). Hence, at leading order in $N$ the variables $z_3^1$ and $z_4^1$ are independent. 
}
\label{fig_test_z_z}
\end{centering}
\end{figure}

\section{Computation of the probability distribution of $z_n^ i $} \label{app2}

Here we compute the probability distribution of $z_n^ i $ at leading order in $N$ . Let us define a super index $L \equiv\{ i_1, \ldots, i_{n-1}\}$, where $L$ stands for `loop', since $L$ represents a loop  passing by the site $i$. Let us also set $X_L \equiv x_{i i_{1}} x_{i_1 i_2}\cdots x_{i_{n-1}i}$.  By Eq. (\ref{5}) one has
\bea \label{61}  
z^ i_n & =  & \frac{\sqrt{(n-1)!}}{N^ {\frac{n-1}{2}}} \sum_{L} X_L, \forall n>2. 
\eea

We observe that the probability distribution of $X_L$ is the same for
every $L$. Hence, according to Eq. (\ref{61}), $z_n^i$ is given by the
sum of equally distributed random variables. Now pick two of these
variables, $X_L, X_{L'}$. For some choices of $L, L'$, $X_{L}$ and
$X_{L'}$ are not independent, since they can depend on the same bond
variables $x_{ij}$. If one picks one variable $X_L$, the number of
variables appearing in the sum (\ref{61}) which are dependent on $X_L$ are those
having at least one common edge with the edges of $X_L$. The number of these variables, at leading order in $N$,  is  $O(N^{n-2})$, since they are obtained by fixing one of the $n-1$ indexes $i_1, \cdots, i_{n-1}$. The latter statement is equivalent to saying that if one picks at random two variables  $X_L, X_{L'}$, the probability that they are correlated is 
\be
O(N^{n-2}/N^{n-1}) = O(N^{-1}). 
\ee 

Hence, at leading order in $N$ we can treat the ensemble of the variables $\{ X_L \} _L $ as independent. According to the central limit theorem, at leading order in $N$ the variable 
\[
 \frac{\sqrt{(n-1)!}}{N^ {\frac{n-1}{2}}}     z^ i_n   =   \frac{1}{\frac{N^ {n-1}}{(n-1)!}} \sum_{L} X_L
\]
is distributed according to a Gaussian distribution with mean $\mathbb{E}_x[X_L]=0$ and variance 
\be \label{var_cl}
\mathbb{E}_x \left[ \left( \frac{\sqrt{(n-1)!}}{N^ {\frac{n-1}{2}}}     z^ i_n \right)^2 \right] = \frac{\mathbb{E}_x[X_L^2]}{\frac{N^ {n-1}}{(n-1)!}} = \frac{1}{\frac{N^ {n-1}}{(n-1)!}},
\ee
where in Eq. (\ref{var_cl}) Eq. (\ref{76}) has been used. It follows that at leading order in $N$, $z_n^i$ is distributed according to a Gaussian distribution with zero mean and unit variance
\be \label{p_of_z}
p_{N}(z_n^i=z) \overset{N \rightarrow \infty}{\rightarrow} \frac{1}{\sqrt{2 \pi}} \textrm{e}^{- \frac{z^2}{2}},  
\ee
where $p_{N}(z_n^i = z)$ is defined as the probability that $z_n^i$ is equal to $z$ at size $N$. 
 \medskip\\

Eq. (\ref{p_of_z}) has been tested numerically for the first few values of $n$:  $p_{N}(z_n^i = z)$ has been computed  by generating  $S \gg 1$  samples of $\{ x \}$, and so of $z_n^i$. For $n=3,4$, the resulting probability distribution $p_{N}(z_n^i = z)$  converges to a Gaussian distribution with zero mean and unit variance as $N$ is increased, confirming the result (\ref{p_of_z}). This convergence is shown in Fig. \ref{fig_test_z}, where $p_{N}(z_4^1 = z)$ is depicted  for different values of $N$ together with the right-hand side of Eq. (\ref{p_of_z}),  as a function of $z$.


\section{Independence of the $z_n^ i$s at leading order in $N$}\label{app3}

Let us consider two distinct variables $z_n^ i, z_m^ j$, and proceed at leading order in $N$.\\

 Following the notation of Appendix \ref{app2}, we write Eq. (\ref{5}) as 
\bea \label{z_i} 
z^ i_n & =&   \frac{\sqrt{(n-1)!}}{N^ {\frac{n-1}{2}}}  \sum_{L} X_L, \\  \label{z_j}
z^ j_m & =  & \frac{\sqrt{(m-1)!}}{N^ {\frac{m-1}{2}}}  \sum_{L'} X_{L'},
\eea
 where $L,L'$ represent a loop of length $n,m$ passing by the site $i,j$ respectively. Some of the variables  $X_L $  depend on some of the variables $X_{L'}$, because they can depend on the same  bond variables $x_{ij}$. Let us pick at random one variable $X_L$ appearing in $z_n^i$, and count the number of variables $X_{L'}$ in $z_m ^j$ that are dependent on $X_L$. At leading order in $N$, these are given by the number of $X_{L'}$ having at least one common bond with $X_L$, and are $O(N^{m-2})$. Hence, if one picks at random two variables $X_L, X_{L'}$ in Eqs. (\ref{z_i}), (\ref{z_j}) respectively, the probability that $X_L, X_{L'}$ are dependent is 
\[
O(N^{m-2}/N^{m-1}) = O (N^{-1}). 
\]

It follows that $z_n^i $ and $z_m^j$ are independent at leading order in $N$, i. e.  for $N \rightarrow \infty$
\be \label{p_z_z}
p_{N}(z_n^i = z, z_m ^j = z') = p_{N}(z_n^i = z) \times p_{N}(z_m ^j = z'), 
\ee
where $p_{N}(z_n^i = z, z_m ^j = z')$ denotes the joint probability that $z_n^i$ equals $z$ and $z_m ^j$ equals $ z'$, at fixed size $N$. \\

Eq. (\ref{p_z_z}) has been tested numerically for $n=3,m=4$:  $p_{N}(z_3^1 = z, z_4^1=z')$ has been computed  by generating  a number $S \gg 1$ of samples of $\{ x \}$, and so of $z_3^1, z_4^1$.  As a result, the left-hand side of Eq. (\ref{p_z_z}) converges to the right-hand side as $N$ is increased, confirming the predictions of the above analytical argument. This is shown in Fig. \ref{fig_test_z_z}, where $p_{1024}(z_3^1 = z, z_4^1=z')$ is depicted together with the $N \rightarrow \infty$-limit of  the right-hand side of Eq. (\ref{p_z_z}) (see Eq. (\ref{p_of_z})),  as a function of $z,z'$.

\bibliographystyle{unsrt}
\bibliography{bibliography}

\end{document}